\documentclass[sigconf,nonacm]{acmart}

\AtBeginDocument{%
  }

\usepackage{algorithm}
\usepackage[noend]{algpseudocode}
\usepackage{array}
\usepackage{multirow}
\usepackage{booktabs}
\usepackage{geometry} 
\usepackage{tabularx} 
\usepackage{caption}  
\usepackage{subcaption}
\usepackage{tcolorbox}
\tcbuselibrary{listings, breakable}
\usepackage{longtable} 
\usepackage{enumitem}  
\usepackage{siunitx} 

\usepackage{listings}
\usepackage{xcolor} 
\lstdefinestyle{mystyle}{
    basicstyle=\ttfamily\small,
    breaklines=true,
    frame=single,
    framerule=0.4pt,
    rulecolor=\color{black!30}, 
    xleftmargin=1em,
}
\newcolumntype{P}[1]{>{\raggedright\arraybackslash}p{#1}}

\newcommand{\appname}{\textsc{FocusGen}}

\usepackage{graphicx}
\usepackage{pifont}
\usepackage[normalem]{ulem}

\begin{document}

\title{{\appname}: Expanding Visual Design Exploration with a Simulated Focus Group of Persona Agents}


\author{Jaewon Choi}
\affiliation{%
  \institution{Hanwha Life}
  \city{Seoul}
  \country{South Korea}}
\email{jaewonch@hanwha.com}

\author{Helena Vasconcelos}
\affiliation{%
  \institution{Stanford University}
  \city{Palo Alto}
  \state{California}
  \country{USA}
}
\email{helenav@stanford.edu}

\author{Hyun Lee}
\affiliation{%
  \institution{KAIST}
  \city{Daejeon}
  \state{}
  \country{South Korea}
}
\email{hyunini@kaist.ac.kr}

\author{Carolyn Zou}
\affiliation{%
  \institution{Stanford University}
  \city{Palo Alto}
  \state{California}
  \country{USA}
}
\email{cqz@stanford.edu}

\author{Tak Yeon Lee}
\affiliation{%
  \institution{KAIST}
  \city{Daejeon}
  \state{}
  \country{South Korea}
}
\email{takyeonlee@kaist.ac.kr}

\author{Michael Bernstein}
\affiliation{%
  \institution{Stanford University}
  \city{Palo Alto}
  \state{California}
  \country{USA}
}
\email{msb@stanford.edu}

\renewcommand{\shortauthors}{Choi et al.}

\begin{abstract}
Creative professionals rarely design for themselves---they design for audiences whose preferences they must anticipate. Yet current text-to-image exploration tools derive diversity entirely from the designer's own input---their prompts, their chosen dimensions, their search queries---confining exploration to what the designer already knows to look for. We present FocusGen, an interactive system that introduces external perspectives into visual design exploration through a ``virtual focus group'' of simulated persona agents. In contrast to prior persona systems in which multiple agents converge as critics on a single evolving artifact, FocusGen uses personas as parallel generators: each agent---constructed from demographic data, a procedurally generated backstory, and aesthetic preferences elicited through interviews---independently drives an iterative generation loop that produces its own visual concept, transforming one design brief into a spectrum of audience-conditioned directions. We evaluate the approach in three parts. With real human participants, we confirm that the iterative refinement loop produces outputs people prefer over zero-shot generation. With synthetic agents at scale, we show that persona conditioning yields higher visual diversity than a generic-assistant baseline---measured by CLIP distance and corroborated by human perceptual judgments---and that open-ended preference interviews yield more diverse outputs than structured ones for both human and synthetic cohorts, while also revealing that agent cohorts recover only part of the diversity of comparable human cohorts. A qualitative study with 16 creative professionals suggests FocusGen helps designers discover unanticipated directions, overcome fixation, and probe audience contexts---while surfacing stereotyping risks that we analyze. We position FocusGen as a divergence scaffold for early-stage ideation rather than a substitute for audience research.
\end{abstract}




\begin{teaserfigure}
  \centering
  \includegraphics[width=\textwidth]{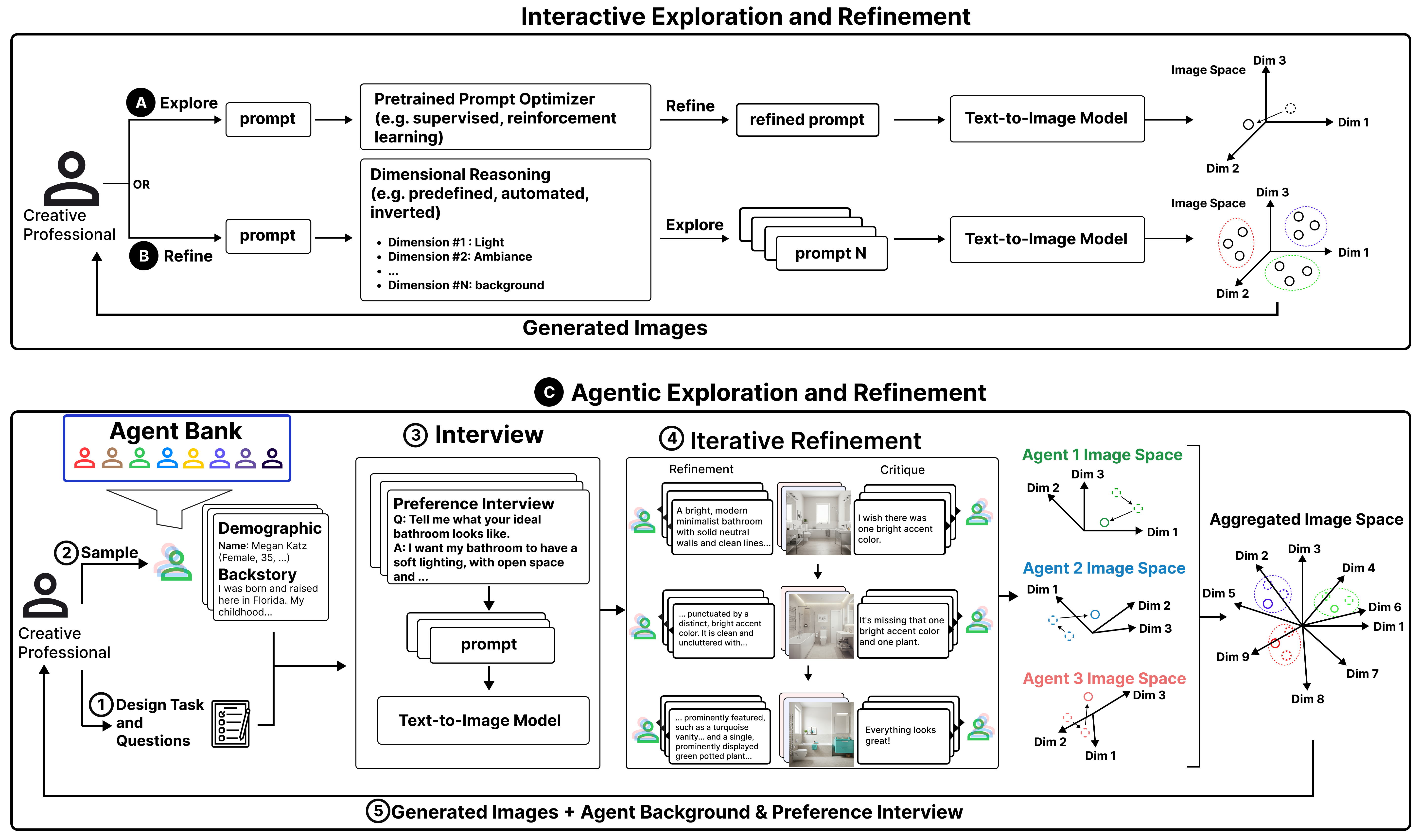}
  \caption{{Comparison of Generative Exploration Paradigms.
  \textbf{(Top) Current Approaches:} (A) Prompt Optimization requires training, which is computationally intensive and difficult to scale.
  (B) Dimensional Reasoning relies on user-defined axes (e.g., Lighting, Ambiance), confining exploration to what the designer already knows to look for.
  \textbf{(Bottom) Our Approach:} (C) {\appname} introduces external perspectives through persona agents equipped with demographics, backstories, and aesthetic preferences. By conducting open-ended \textit{preference interviews}, the system elicits dimensions the designer may not have considered. Each agent independently generates visual concepts conditioned on its distinct profile, surfacing diverse directions---the designer's blind spots---to support creative professionals in divergent exploration.}}
  \label{fig:teaser}
\end{teaserfigure}

\maketitle

\section{Introduction}
Creative professionals rarely design in a vacuum. Whether crafting a brand identity, a product concept, or a marketing campaign, their work must resonate with audiences whose preferences differ from their own. In the early stages of design, exploring diverse visual directions is critical---yet this is precisely where current text-to-image tools fall short. Not because the models lack capability, but because the exploration process is fundamentally bounded by the designer's own perspective.

Today's exploration tools all share a common limitation: they derive diversity from the \textit{user's} input. Prompt engineering~\cite{oppenlaender_creativity_2022, xie_prompt_2023} relies on the designer to articulate the right modifiers. Dimensional reasoning tools like Luminate~\cite{suh_luminate_2024} and DesignAID~\cite{cai_designaid_2023} help users navigate semantic axes, but those axes are extracted from the user's own prompts or generated images. Even automated prompt optimization~\cite{hao_optimizing_2023, wang_promptcharm_2024} refines what the user has already expressed. These approaches can surface ``known unknowns''---variations the designer suspects exist but hasn't yet explored---but they cannot surface what the designer has not thought to look for: the \textit{unknown unknowns}.

In professional practice, these unknown unknowns are traditionally surfaced through audience feedback---focus groups, user interviews, and design critiques that bring external perspectives into the process. But gathering such feedback is expensive, slow, and impractical during early-stage ideation when concepts are still forming. We ask: can simulated audience perspectives serve this role?

We present \textbf{\appname}, an interactive system that introduces external perspectives into visual design exploration through a ``virtual focus group'' of persona agents. Prior work has shown that LLM-based agents grounded in demographics and rich backstories can produce idiosyncratic, individuated responses rather than generic stereotypes~\cite{park_generative_2023, park_generative_2024, moon_virtual_2024}. {\appname} builds on this foundation by adding a layer of \textbf{aesthetic preference elicitation}: each agent is interviewed to surface its specific visual priorities, grounded in its life context. Each agent's elicited preferences are supplied to an iterative generation and critique loop. This workflow is intended to make alternatives interpretable through stated persona preferences. Our evaluations test whether heterogeneous persona-profile conditioning increases output diversity, but they do not establish that outputs are discriminably faithful to individual personas or representative of demographic groups.


{\appname} constructs its agents synthetically: demographic profiles are sampled from census data~\cite{castricato_persona_2024}, enriched with procedurally generated backstories~\cite{moon_virtual_2024}, and grounded through open-ended interviews that the agents answer in character. When given a design brief, each agent independently generates visual concepts conditioned on its distinct profile. The system parallelizes this across the full agent cohort, producing a gallery of audience-specific visual concepts. To validate this approach, we first confirmed with real human participants that the refinement loop produces outputs people prefer over zero-shot generation, then showed with synthetic agents at scale that persona diversity translates into measurable visual diversity.

In summary, the contributions of this work are:
\begin{itemize}
    \item \textbf{{\appname}:} A parallelized-divergent interaction model for persona-based creativity support---persona agents as independent, parallel generators of audience-conditioned design alternatives, rather than convergent critics of a single evolving artifact---instantiated in {\appname} with a 1,000-agent, demographically grounded Agent Bank and an interface that shifts the designer's role from prompt engineering to audience curation.
    \item \textbf{Empirical evidence} that (i) the iterative refinement loop produces outputs that human participants prefer over zero-shot generation, (ii) persona conditioning measurably and perceptibly increases the visual diversity of generated alternatives over a generic-assistant baseline, and (iii) open-ended preference interviews expand output diversity relative to structured interviews for both human and synthetic cohorts---together with an explicit characterization of what these evaluations do not establish (discriminative persona faithfulness and demographic representativeness) and of a persistent human--agent diversity gap.
    \item \textbf{The design insight} that preference interviews---where designers craft questions for simulated personas rather than prompts for a model---serve as a productive interaction paradigm for divergent exploration, shifting the designer's role from commanding to curating, with qualitative evidence from 16 creative professionals on how this paradigm functions in practice.
\end{itemize}


\section{Related Work}

\subsection{Designing for Audiences: The Need for External Perspectives}

Professional designers rarely design for themselves---they design for audiences whose preferences, contexts, and reactions they must anticipate. This is well-established in the design literature: designers are inherently not their users, and deliberate methods are needed to bridge this empathy gap~\cite{kouprie_framework_2009, bennett_promise_2019}. In practice, external feedback---from focus groups, user interviews, and design critiques---provides genuinely different information than what designers generate internally~\cite{luther_structuring_2015}. Systems like Voyant~\cite{xu_voyant_2014} demonstrated that even non-expert crowd workers, when selected to match a target audience demographic, can provide perception-oriented feedback that meaningfully changes design outcomes. Furthermore, creating and comparing multiple design alternatives in parallel leads to better results and greater divergence than sequential refinement~\cite{dow_parallel_2010}---a principle {\appname} enacts by parallelizing feedback across a diverse agent population.

However, gathering real audience feedback is expensive, slow, and impractical during early-stage ideation when concepts are still forming. This motivates the core question of our work: can \textit{simulated} audience perspectives serve this role?

\subsection{Current Exploration Tools Confine Discovery to Known Unknowns}

\subsubsection{Design Fixation and AI-Induced Homogenization}
Design fixation---the unconscious adherence to a limited set of ideas that restricts exploration~\cite{jansson_design_1991, crilly_where_2017}---is a well-documented challenge that affects novices and experts alike~\cite{cross_expertise_2004}. Recent work has shown that generative AI tools can \textit{exacerbate} rather than mitigate this problem. Wadinambiarachchi et al.~\cite{wadinambiarachchi_effects_2024} found that AI image generators increase fixation compared to traditional inspiration sources, producing fewer ideas with less variety. At the collective level, Doshi and Hauser~\cite{doshi_generative_2024} demonstrated that while LLM access improves individual creative output, it significantly reduces the \textit{diversity} of ideas across users---a homogenization effect confirmed across multiple LLMs~\cite{jakesch_co-writing_2023}. This creates a paradox: tools designed to support creativity may narrow the very design space they aim to expand.

\subsubsection{Prompt-Based Exploration and Its Limits}
The primary means of controlling text-to-image models remains the text prompt, and prompt engineering is difficult~\cite{oppenlaender_creativity_2022, xie_prompt_2023}. Users explore prompts opportunistically rather than systematically, constrained by what they can articulate and imagine~\cite{zamfirescu-pereira_why_2023}. Subramonyam et al.~\cite{subramonyam_bridging_2024} formalized this as the ``Gulf of Envisioning''---users can only prompt for what they can envision, creating a cognitive bottleneck that bounds exploration to \textit{known unknowns}.

Automated prompt optimization~\cite{hao_optimizing_2023, wang_promptcharm_2024} and iterative refinement frameworks like Self-Refine~\cite{madaan_self-refine_2023} and reflexion~\cite{shinn_reflexion_2023} reduce this burden by using LLMs to critique and revise outputs. However, these approaches still optimize for the user's stated intent using a generic critic---not the preferences of a specific audience.

\subsubsection{Dimensional Reasoning Tools}
Recent creativity support tools facilitate exploration through \textit{dimensional reasoning}---navigating the design space via semantic axes. Luminate~\cite{suh_luminate_2024} generates dimensions for users to adjust, DesignAID~\cite{cai_designaid_2023} generates diverse ideas from user text, DesignWeaver~\cite{tao_designweaver_2025} extracts dimensions from design documents, and POET~\cite{han_poet_2025} discovers dimensions of homogeneity via prompt inversion. While valuable, these tools share a fundamental limitation: their dimensions are derived from the user's own input, confining exploration within the user's existing conceptual frame. They help designers explore ``known unknowns'' but cannot surface the \textit{unknown unknowns}---perspectives the designer has not thought to look for~\cite{jensen_eliciting_2017}.

{\appname} addresses this gap by introducing \textit{external} perspectives into the generation loop. Unlike prompt recommendation systems~\cite{brade_promptify_2023, son_genquery_2024, feng_promptmagician_2023} that optimize for the user's predicted intent, we simulate a diverse audience whose feedback may contradict or expand the designer's initial framing, supporting divergent ideation rather than convergence.

\subsection{Simulating Populations with Persona Agents}

The development of generative agents that simulate specific people offers a path toward scalable audience-like feedback. Park et al.~\cite{park_generative_2023} demonstrated that LLMs can generate believable, individualized behaviors, and Social Simulacra~\cite{park_social_2022} showed that designers used simulated community members to see ``beyond the social interactions that they intend their design to produce.'' These agents can be grounded in demographic data~\cite{castricato_persona_2024, argyle_out_2023}, rich narrative contexts~\cite{moon_virtual_2024}, or specific user preferences~\cite{choi_proxona_2025, ryan_synthesizeme_2025}, maintaining character consistency across tasks~\cite{tseng_two_2024, shao_character-llm_2023}.

However, the reliance on demographic conditioning necessitates caution regarding \textit{representational harms}~\cite{weidinger_taxonomy_2022, li_actions_2025}. Park et al.~\cite{park_generative_2024} demonstrate that enriching agents with specific life histories substantially mitigates stereotyping by shifting the model from broad generalizations to individual traits.

Several recent systems apply persona agents to creative and intellectual work, and they do more than post-hoc critique. PosterMate~\cite{shin_postermate_2025} lets audience persona agents propose edits to text, imagery, and theme that are integrated directly into the poster canvas; Proxona~\cite{choi_proxona_2025} supports creators in making sense of their audience and refining early-stage creative work through conversations with data-grounded audience personas; PersonaFlow~\cite{liu_personaflow_2025} uses simulated domain experts to drive literature retrieval, critique, and the generation of revised research questions; and SimTube~\cite{hung_simtube_2024} generates simulated audience commentary on video. We therefore do not position {\appname} as the first system to bring personas into a generative loop. The distinction we draw is one of interaction topology. These systems are predominantly \textit{convergent}: multiple personas critique, annotate, or refine a single evolving artifact that the creator owns. {\appname} is \textit{parallel-divergent}: each persona independently authors its own visual concept end-to-end---preference elicitation, prompt construction, and iterative refinement---and the system's unit of output is the resulting spectrum of audience-conditioned artifacts rather than an improved single artifact. This topological choice is motivated by evidence that developing alternatives in parallel produces more divergent outcomes than sequentially refining one concept~\cite{dow_parallel_2010}; parallelizing across personas is what allows population-level diversity to translate into visual diversity, the property we test in Evaluations 1 and 2.

A complementary line of work uses multi-agent coordination to improve generative image creation itself. CREA~\cite{CREA} orchestrates agents in creative-process roles---such as a creative director, critics, and editors---that collaborate on editing and refining images. The axis of agent differentiation differs from ours: CREA's agents embody stages of the creative process and converge on one high-quality output, whereas {\appname}'s agents embody members of a simulated audience and diverge into many outputs. The two approaches are composable---CREA-style role specialization could strengthen the per-agent refinement loop inside each {\appname} persona---but they target different problems: making one image better versus exposing a designer to directions they did not think to look for.

{\appname}'s individual components---demographic persona construction~\cite{castricato_persona_2024}, narrative grounding~\cite{moon_virtual_2024}, and iterative self-refinement~\cite{madaan_self-refine_2023, shinn_reflexion_2023}---each build on established techniques. The contribution lies in their composition into the parallelized-divergent workflow above: using each persona's elicited preferences to drive independent generation, parallelizing across a demographically diverse agent population, and providing an interface that lets designers act as curators of a simulated audience. Whether this translation of population diversity into visual diversity actually occurs is an empirical question, which we test in Section 4.

\section{{\appname}: System Design}
\label{sec:method}

This section describes {\appname}, an interactive system that introduces synthetic external audience perspectives into text-to-image design exploration. We first describe the designer-facing workflow---how a designer configures, launches, and explores results---then detail the two underlying mechanisms that power it: persona initialization and iterative image generation.

\paragraph{Implementation Details.}
For all experiments and system instantiations described in this paper, {\appname} was implemented using Google's \textit{gemini-2.5-flash} Large Language Model for agent simulation and critique, and Google's \textit{imagen-3.0-generate} text-to-image model for visual generation.

\begin{figure*}[ht]
    \centering
    \includegraphics[width=0.8\textwidth]{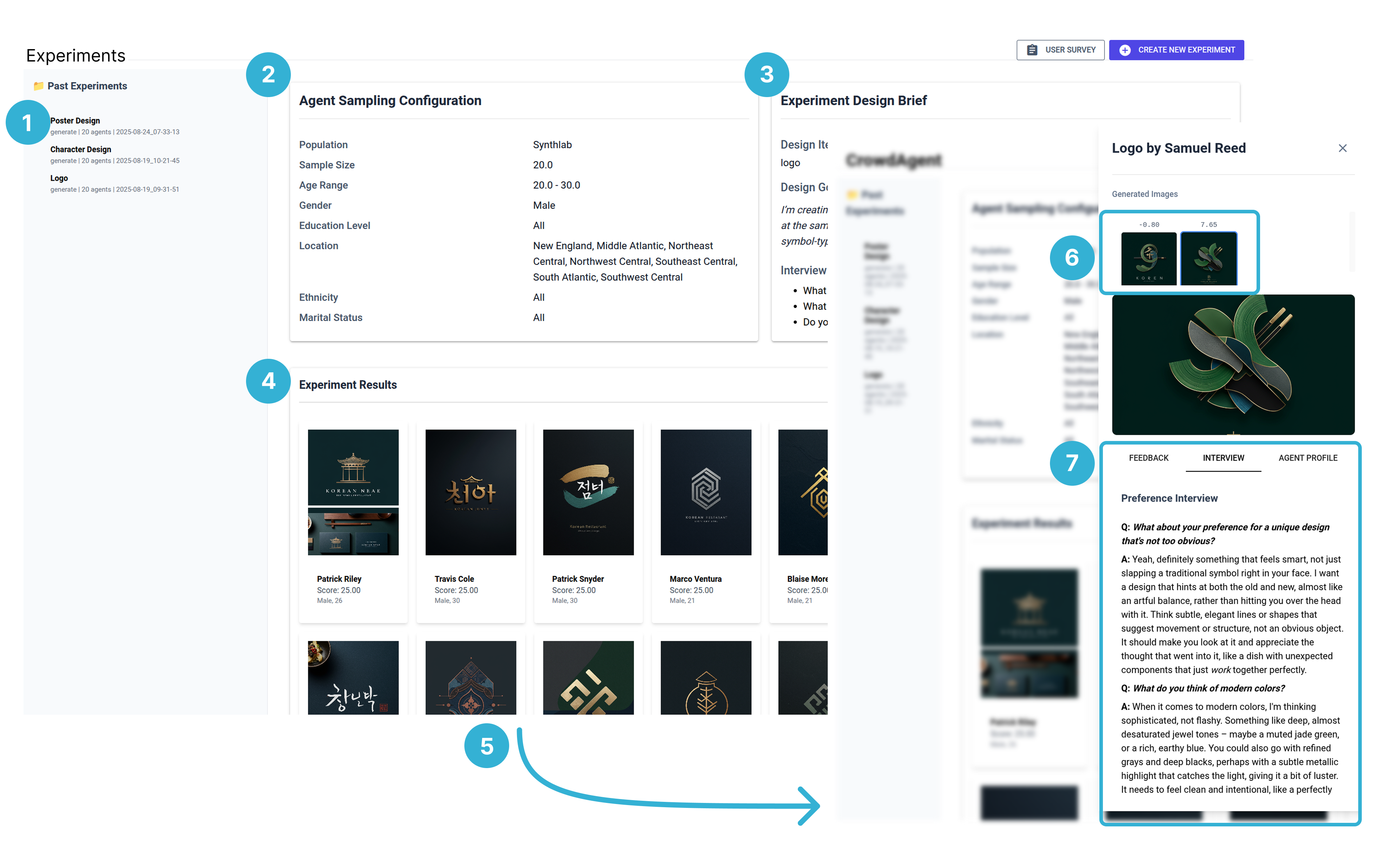}
    \caption{The \textbf{Feedback Exploration View}. \textbf{(1)} The navigation panel lists all past experiments for easy access. When an experiment is selected, its \textbf{(2)} agent configuration and \textbf{(3)} design brief are displayed. \textbf{(4)} The main gallery populates with a mood board of images from the persona crowd.  \textbf{(5)} Clicking an image opens a detail view, revealing \textbf{(6)} the iterative refinement history, \textbf{(7)} the agent's full textual feedback and interview responses, and its demographic profile.}
    \Description{A screenshot of the 'Feedback Exploration View,' annotated with seven numbered labels. Label 1 points to the 'Past Experiments' navigation panel. Labels 2 and 3 indicate the summary panels for the agent configuration and design brief. Label 4 shows the main gallery of final generated images. An arrow, labeled 5, shows that clicking an image opens a detail view. Within this view, label 6 points to the iterative history as a series of thumbnails, and label 7 points to the detailed textual information, including the agent's interview responses.}
    \label{fig:feedback-exploration-workflow}
\end{figure*}

\subsection{Designer Workflow}

{\appname} provides designers with access to a diverse ``Agent Bank''---a repository of 1000 persona agents, each initialized with demographic data~\cite{castricato_persona_2024} and a procedurally generated backstory~\cite{moon_virtual_2024}. Because each agent brings a distinct perspective grounded in its unique life context, the system can transform a single design brief into a gallery of audience-specific visual directions. The designer curates this process through a two-stage workflow.

\subsubsection{Stage 1: Task Creation}
The designer first selects a cohort of agents using demographic filters (age, gender, education, income) to target specific audience segments. They then define the creative brief: a high-level \textbf{Design Objective} and a set of \textbf{preference interview questions} (see Table~\ref{tab:interior_design_questions}). The interview step is critical---it requires designers to shift from \textit{commanding} an AI to \textit{interviewing} a persona. Questions can range from specific and orthogonal (e.g., separating ``lighting'' from ``materials'') to broad and open-ended (e.g., ``Describe your five most important preferences''). As we show in Evaluation 2, this choice significantly affects the diversity of the resulting outputs. Once configured, the system generates visual concepts for each agent in parallel.

\subsubsection{Stage 2: Feedback Exploration}
The \textbf{Feedback Exploration View} (Figure~\ref{fig:feedback-exploration-workflow}) displays a gallery of the final images from all agents, each tagged with the generating agent's demographics. Clicking any image reveals a \textbf{detail view} showing the full iterative refinement history (with preference scores per iteration), the agent's structured critiques, interview responses, and complete profile. This allows designers to understand not only \textit{what} was generated but \textit{why}, connecting visual output to persona reasoning. Insights from one round inform subsequent experiments with refined filters, goals, or interview questions.

\subsection{Persona Initialization}

Behind the scenes, each agent in the Agent Bank is constructed through a two-layer process that establishes both a stable identity and task-specific aesthetic preferences.

\subsubsection{Layer 1: Core Identity}
The core identity combines structured demographic data (e.g., age, ethnicity, occupation) with a rich personal backstory (a first-person narrative detailing life events, hobbies, and memories). Demographic profiles are sampled from real-world datasets like the U.S. Census~\cite{castricato_persona_2024}, and the backing LLM procedurally generates a unique, coherent backstory for each profile~\cite{moon_virtual_2024}. This procedural approach enables {\appname} to construct the Agent Bank at scale without requiring real participant data. To validate the profile-to-image stage of this pipeline, we first tested it with preference profiles collected from real human participants (Technical Validation, Section 4.1) before applying it to synthetic agents (Evaluations 1 and 2).

\subsubsection{Layer 2: Preference Elicitation}
When the designer launches a task, each selected agent is prompted to answer the interview questions in character, inferring preferences from its demographic profile and backstory (e.g., an agent whose backstory centers on carpentry might cite ``natural timber finishes''); we treat such responses as the agent's operative preferences for generation, without claiming they mirror the preferences of real members of the corresponding group (Section 6.4). The output is a standardized textual \textit{Persona Profile} that feeds into the image generation pipeline. Concretely, the agent prompt concatenates four components: the demographic profile ($\approx$150 tokens), a procedurally generated backstory ($\approx$1000 tokens), the elicited preference interview responses ($\approx$400 tokens), and a roleplay instruction ($\approx$200 tokens). A full example is provided in the Appendix (Table~\ref{tab:agent-prompts}).

\subsection{Iterative Image Generation}

The final step translates each agent's Persona Profile into a visual concept aligned with that profile. Because text-to-image models often miss details in complex prompts~\cite{agrawala_unpredictable_2023}, a single-shot generation may not fully capture a persona's preferences. The system therefore employs an iterative refinement loop~\cite{madaan_self-refine_2023, shinn_reflexion_2023}.

At each iteration, the agent critiques the generated image against its Persona Profile, identifying elements to preserve and deviations to correct, then generates a revised prompt. The agent maintains a history of past prompts and critiques to diagnose persistent failures. The loop terminates when no deviations remain or a maximum of $N=5$ iterations is reached. Because the process is stochastic and non-monotonic, the agent reviews the entire image sequence and selects the best via pairwise TrueSkill ranking~\cite{herbrich_trueskill_2006}.

\begin{table}[t]
\centering
\small
\begin{tabular}{l p{5.8cm}}
\toprule
\multicolumn{2}{l}{\textbf{Goal:} Design the ideal bedroom.} \\
\midrule
\multicolumn{2}{l}{\textit{Format A: Structured Questions}} \\
\midrule
1 & Ideal colors and materials for walls, floor, ceiling. \\
2 & Ideal bed (shape, frame, bedding colors, textures). \\
3 & Other furniture or items to include. \\
4 & Type of lighting (ceiling, bedside, accent). \\
5 & Window coverings (curtains, blinds, material). \\
\midrule
\multicolumn{2}{l}{\textit{Format B: Open-Ended Question}} \\
\midrule
1 & Describe five or more preferences for your bedroom in order of importance. \\
\bottomrule
\end{tabular}
\caption{Example interview questions for a bedroom design task. Designers craft their own questions, which can range from specific (Format A) to open-ended (Format B). This choice significantly affects output diversity (Evaluation 2).}\Description{A compact table showing two interview formats for a bedroom design task. Format A lists five specific questions about colors, bed, furniture, lighting, and windows. Format B uses a single open-ended prompt asking for five
preferences in order of importance.}
\label{tab:interior_design_questions}
\end{table}

\section{Evaluations}
Our evaluations are designed to support three claims, and we are explicit about a fourth that they do not support. They show (1) that, given a fixed preference profile grounded in a real person's data, the iterative refinement loop produces outputs that the person prefers over zero-shot generation (Technical Validation, Section 4.1); (2) that conditioning generation on diverse personas yields measurably and perceptibly more diverse output sets than a generic-assistant baseline (Evaluation 1, Section 4.2); and (3) that open-ended preference elicitation yields more diverse outputs than structured elicitation, a pattern that holds for both human participants and synthetic agents (Evaluation 2, Section 4.3). They do not show (4) that a synthetic agent's outputs are \textit{discriminably faithful} to its specific persona---for example, that a participant would prefer images generated from their own profile over images generated from someone else's---nor that synthetic agents' self-generated preferences are representative of the demographic groups they are sampled from. Claim (4) requires a discriminative test that was not part of this work; we describe it in Section 6.4 and treat it as the most important next validation step. The qualitative study (Section 5) examines how creative professionals use the system in practice, independent of claim (4).

All studies were approved by the Institutional Review Board (IRB) at the institution where the study occurred.

\subsection{Technical Validation: Iterative Refinement Produces More Preferred Outputs}

Before testing whether diverse personas produce diverse outputs, we validate the generation pipeline itself: given a preference profile elicited from a real person, does the iterative loop produce images that the person actually prefers? We emphasize the scope of this validation. It tests the profile-to-image stage of the pipeline, holding the preference profile fixed; it does not test the persona-to-preference stage (whether synthetic agents generate preferences representative of the people or groups they are modeled on), and, because it lacks a generic-critic control and a cross-persona comparison, it cannot by itself distinguish persona-specific alignment from generic image improvement. We recruited 27 participants from Prolific, collected their demographic profiles, backstories, and aesthetic preferences, and used this data to construct persona agents that served as digital proxies for each participant.

\subsubsection{Method}
Each session proceeded in three stages. First, the participant's data was used to initialize a persona agent. Second, the system generated a sequence of up to five candidate images using the iterative refinement pipeline, terminating early if the agent identified no deviations. Third, both the participant and their corresponding agent independently ranked the full image sequence through pairwise comparisons (without knowledge of generation order), yielding TrueSkill scores~\cite{herbrich_trueskill_2006} that track perceived quality across iterations.

\subsubsection{Results}

\paragraph{Users Actively Prefer Refined Images.}
Participants selected a refined image (i.e., any image beyond the initial zero-feedback generation) as their top choice in the majority of sessions. To confirm this preference was not merely an artifact of there being more refined images (up to four) than initial images (one) in each session, we compared the observed choices to a random-selection baseline adjusted for this imbalance. A chi-squared goodness-of-fit test confirmed that users selected refined images significantly more often than this adjusted baseline ($\chi^2(1, N=106) = 5.20, p < .05$), demonstrating that the preference is statistically robust and not simply a product of the number of options presented.

\paragraph{The Final Outcome is a Marked Improvement over the Initial Image.}
This preference is supported by a significant increase in alignment scores. A paired samples t-test confirmed that the average TrueSkill score of users' final chosen image ($M = 15.57$) was more than double that of their initial, zero-feedback image ($M = 7.30$, $p < .001$). This large effect size demonstrates that the iterative refinement pipeline produces a meaningfully better final outcome.

\paragraph{Human--Agent Ranking Agreement.}
We also examined whether human and agent TrueSkill rankings of the full image sequence were correlated. The correlation was weak ($r = .26$, $p =.19$). Plausible contributors include differences in evaluation style---human evaluators judge holistically, weighing composition, mood, and coherence, while the agent critiques literal adherence to stated preferences---and the stochastic, non-monotonic nature of text-to-image refinement. We treat this result as a substantive caveat rather than an incidental one: a weak correlation means the agent's judgment of ``best for this profile'' only partially tracks the profile owner's own judgment. The preceding results therefore establish that refinement produces outcomes people prefer; they do not establish that an agent's selections are discriminably aligned with its specific persona. The direct test---asking participants to choose between outputs generated from their own profile and outputs generated from another participant's profile, and including a generic-critic refinement arm---was not part of this study and is the most important next validation step (Section 6.4).

\subsection{Evaluation 1: Agent Bank Diversity Translates to Visual Diversity}
Having established that participants generally prefer refined outputs over the initial generation (Section 4.1), we now test whether the demographic and narrative diversity of our simulated agents translates into greater visual diversity. Using the Agent Bank that we constructed (U.S. census demographics, procedural backstories~\cite{castricato_persona_2024, moon_virtual_2024}), we compared the collective output from a diverse agent cohort against outputs produced without personalization.

\paragraph{Conditions.}
For each task, we generated $3 \times 3$ grids of nine images under two conditions. In the \textbf{Agent-Enabled} condition, each image was generated by a unique Persona Agent from the Agent Bank. In the \textbf{Agent-Disabled} baseline, the persona layer was deactivated and a generic LLM assistant (prompt: ``You are a helpful and creative assistant that can suggest new and inspiring designs. Given the task, think outside the box and provide imaginative and innovative suggestions'') served as the critic. Both conditions used the same iterative refinement process. We quantify diversity using average pairwise CLIP distance and CLIP dispersion. We note that CLIP distance measures semantic and visual dissimilarity but not design quality or relevance---random, off-brief images would also score high. We therefore complement it with a human perceptual evaluation.

Both conditions received the identical design brief and the identical refinement scaffold (the same critique-and-revise prompt template, maximum of $N=5$ iterations, and TrueSkill selection step). The only difference was the identity supplied to the critic: in the Agent-Enabled condition, the critic received the full persona profile (demographics, backstory, elicited preferences, roleplay instruction; see Table~\ref{tab:agent-prompts}); in the Agent-Disabled condition, it received only the generic-assistant prompt above.

\paragraph{Procedure.}
We recruited 20 participants from Prolific who viewed randomized pairs of grids side-by-side across five creative tasks and selected the more visually diverse one (or `Tie'; see Figure~\ref{fig:persona_agent_evaluation_interface} in the Appendix).

\subsubsection{Results}
The \textbf{Agent-Enabled} condition yielded higher CLIP diversity across all five tasks (Table~\ref{tab:clip_persona}): \textbf{58\%} average increase in pairwise distance and \textbf{33\%} in dispersion, with the most dramatic gain in Interior Design (\textbf{180\%}), likely because generic models produce especially homogeneous outputs in this domain (defaulting to modern minimalist styles). Participants also significantly preferred the Agent-Enabled grids as more visually diverse ($\chi^2(2, N=468) = 86.81, p < .001$). Together, these results show that persona-profile conditioning produced grids with higher CLIP diversity that participants also perceived as more diverse than grids generated using a shared generic critic.

\paragraph{Limitations of This Comparison.}
The persona condition introduced substantially more and more heterogeneous conditioning text than the generic-critic condition, including demographic information, a procedurally generated backstory, elicited preference responses, and a roleplay instruction. This comparison therefore establishes a system-level diversity gain from the full persona-profile conditioning pipeline, but does not isolate which profile components caused that gain. In particular, the observed difference may reflect persona semantics, prompt length, lexical heterogeneity, preference variation, or a combination of these factors.

\subsection{Evaluation 2: Open-Ended Interviews Expand Design Diversity}

We now test the claim: that the way preferences are elicited affects the diversity of the resulting visual concepts.

\subsubsection{Methodology}
We compared Structured interviews against Open-Ended interviews across five creative tasks, evaluating both a \textbf{Human Cohort} (40 Prolific participants, $N_{open}$=23, $N_{struct}$=17) and an \textbf{Agent Cohort} (24 synthetic agents from the Agent Bank, interviewed using both formats). In the Structured condition, designers provide specific, orthogonal questions (e.g., separating ``lighting'' from ``materials''). In the Open-Ended condition, agents receive a single broad prompt (e.g., ``Describe your five most important preferences''), allowing them to self-define the dimensions that matter most. We quantify diversity using average pairwise CLIP distance and CLIP dispersion.

\begin{table}[t]
\centering
\small
\caption{Open-ended interviews produce higher visual diversity for both humans and synthetic agents. Values show CLIP Distance with Dispersion in parentheses.}
\Description{A table comparing CLIP Distance and Dispersion for Structured versus Open-Ended interviews across five tasks for Human and Agent cohorts.}
\label{tab:clip_interview}
\begin{tabular}{l cccc}
\toprule
& \multicolumn{2}{c}{\textbf{Humans}}
& \multicolumn{2}{c}{\textbf{Agents}} \\
\cmidrule(lr){2-3} \cmidrule(lr){4-5}
\textbf{Task} & {Struct.} & {\textbf{Open}}
              & {Struct.} & {\textbf{Open}} \\
\midrule
UI Design  & .27\,(.10) & \textbf{.34}\,(\textbf{.14})
           & .24\,(.10) & \textbf{.29}\,(\textbf{.12}) \\
Packaging  & .20\,(.09) & \textbf{.26}\,(\textbf{.11})
           & .21\,(.09) & \textbf{.24}\,(\textbf{.10}) \\
Logo       & .20\,(.09) & \textbf{.25}\,(\textbf{.11})
           & .09\,(.03) & \textbf{.12}\,(\textbf{.05}) \\
Interior   & .20\,(.08) & \textbf{.25}\,(\textbf{.10})
           & .10\,(.04) & \textbf{.14}\,(\textbf{.06}) \\
Food       & .13\,(.05) & \textbf{.21}\,(\textbf{.10})
           & .09\,(.03) & \textbf{.14}\,(\textbf{.06}) \\
\midrule
\textbf{Avg} & .20\,(.08) & \textbf{.26}\,(\textbf{.10})
             & .15\,(.06) & \textbf{.19}\,(\textbf{.08}) \\
\bottomrule
\end{tabular}
\end{table}

\subsubsection{Results}
Results from 115 user sessions and 120 agent sessions (Table~\ref{tab:clip_interview}) confirm that Open-Ended interviews produce higher diversity in both cohorts. For humans, Open-Ended increased CLIP Distance by \textbf{30\%} and Dispersion by \textbf{25\%}. For synthetic agents, the pattern was consistent (\textbf{27\%} and \textbf{33\%} increases, respectively). Notably, the Open-Ended condition produced higher CLIP Distance in all five tasks for both cohorts. A sign test confirms this perfect consistency is unlikely by chance ($p = .031$, one-sided), though we note this test has minimal statistical power with only five data points---the result is driven entirely by the consistency of the pattern rather than the magnitude of individual effects. We therefore treat this as a \textit{consistency check} rather than our primary evidence.

\paragraph{The Human-Agent Diversity Gap.}
Comparing the cohorts reveals a limitation we want to state precisely. Agents mirror the direction of the human effect (open-ended elicitation increases diversity in both cohorts), but their absolute diversity is consistently lower, and on Logo, Interior, and Food the gap is roughly a factor of two (e.g., Food, open-ended: humans .21 vs. agents .14). This bounds the ``virtual focus group'' framing in two ways. First, it bounds the claim: {\appname} expands exploration relative to a designer's solo prompting and relative to a generic assistant (Evaluation 1), but it does not reproduce the variability that a comparable human cohort would supply. A designer who treats the agent gallery as a census of audience reactions would be systematically under-exposed to the tails of real preference distributions. Second, it suggests a mechanism. LLM outputs are known to homogenize across users and prompts~\cite{doshi_generative_2024, jakesch_co-writing_2023}, and all of our agents share a single base model; demographic profiles and procedural backstories shift that model's outputs but evidently do not recover the idiosyncrasy of real individuals. Consistent with this account, the gap is smallest on tasks with strong conventional structure (UI design, packaging---where agents approach human diversity) and largest where personal taste dominates (logos, interiors, food). The practical implication is that {\appname} should be used as a divergence scaffold---a fast first pass that surfaces directions worth testing with real audiences---rather than as a substitute for audience research, and we have calibrated our claims throughout the paper accordingly. Closing the gap, for example via heterogeneous base models or personas grounded in real interview data~\cite{park_generative_2024}, is a primary direction for future work.

\paragraph{Limitations of This Comparison.}
We acknowledge that the Structured and Open-Ended conditions differ in both question \textit{format} and \textit{degree of constraint} (five prescribed dimensions vs.\ agent-defined dimensions), so we cannot isolate which factor drives the diversity gains. Nonetheless, the finding provides actionable guidance: when the goal is divergent exploration, open-ended questions that let agents self-define relevant dimensions produce more diverse outputs.

\begin{table}[t]
\centering
\small
\caption{Persona agents produce higher visual diversity than generic baselines across all tasks. Values show CLIP Distance with Dispersion in parentheses. $\Delta$ is computed from the two-decimal values shown.}
\Description{A table comparing CLIP Distance and Dispersion for Generic versus Persona agents across five tasks.}
\label{tab:clip_persona}
\begin{tabular}{l ccc}
\toprule
\textbf{Task} & {\textbf{Generic}} & {\textbf{Persona}}
              & {\textbf{$\Delta$(\%)}} \\
\midrule
UI Design  & .22\,(.11) & \textbf{.29}\,(\textbf{.12}) & +32\%\,(+9\%) \\
Packaging  & .18\,(.09) & \textbf{.24}\,(\textbf{.10}) & +33\%\,(+11\%) \\
Logo       & .08\,(.03) & \textbf{.12}\,(\textbf{.05}) & +50\%\,(+67\%) \\
Interior   & .05\,(.02) & \textbf{.14}\,(\textbf{.06}) & +180\%\,(+200\%) \\
Food       & .07\,(.03) & \textbf{.14}\,(\textbf{.06}) & +100\%\,(+100\%) \\
\midrule
\textbf{Avg} & .12\,(.06) & \textbf{.19}\,(\textbf{.08}) & \textbf{+58\%}\,(\textbf{+33\%}) \\
\bottomrule
\end{tabular}
\end{table}

\section{User Study: {\appname} in Professional Workflows}
We conducted an exploratory qualitative study with 16 creative professionals to understand how the persona-driven, parallel-divergent workflow functions in professional ideation practice. Each session comprised two parts: a grounding task in which participants worked on a brief with their own preferred generative AI tool, followed by the same brief with {\appname} (populated with U.S. census demographics~\cite{castricato_persona_2024} and procedural backstories). The first session was designed to ground participants in a concrete instance of their normal workflow that they could articulate against---not to serve as a controlled comparison condition (see Study Design Rationale below). We designed the study to answer:

\begin{itemize}
    \item[\textbf{RQ1:}] How do creative professionals experience and evaluate agent-driven parallel ideation---what value and what limitations do they perceive in it within their workflows?
    \item[\textbf{RQ2:}] In what ways do creative professionals utilize the diverse agents and their associated preferences to explore a design space and generate novel ideas?
\end{itemize}

\subsection{Participants}
We recruited 16 creative professionals (13 female, 3 male; avg. 7.75 years of experience) through online ads and personal contacts. All had professional experience with generative AI tools. The most commonly used tools were ChatGPT and Midjourney. A detailed summary is in Table~\ref{tab:participants_info} in the Appendix.

\subsection{Study Task}
Participants chose one of six creative ideation tasks (e.g., logo design for a fintech company, poster for a coffee festival; see Table~\ref{tab:design_tasks_overview} in the Appendix). They were asked to produce 3-5 distinct visual concepts with reference images and rationales, mimicking early-stage creative workflows.

\subsection{Procedure}
The study was conducted remotely via Zoom ($\sim$2 hours per session, audio/video recorded with IRB approval). Participants completed two tasks in fixed order:

\paragraph{Task 1: Baseline.} Participants selected a task adjacent to (but distinct from) their core expertise and performed ideation using their preferred generative AI tool, compiling 3-5 concepts in a shared Google Doc.

\paragraph{Task 2: {\appname}.} After a 15-minute tutorial framing the system as a ``virtual focus group,'' participants performed the same task using {\appname} and compiled their findings.

\begin{figure*}[tb]
    \centering
    \includegraphics[width=\textwidth]{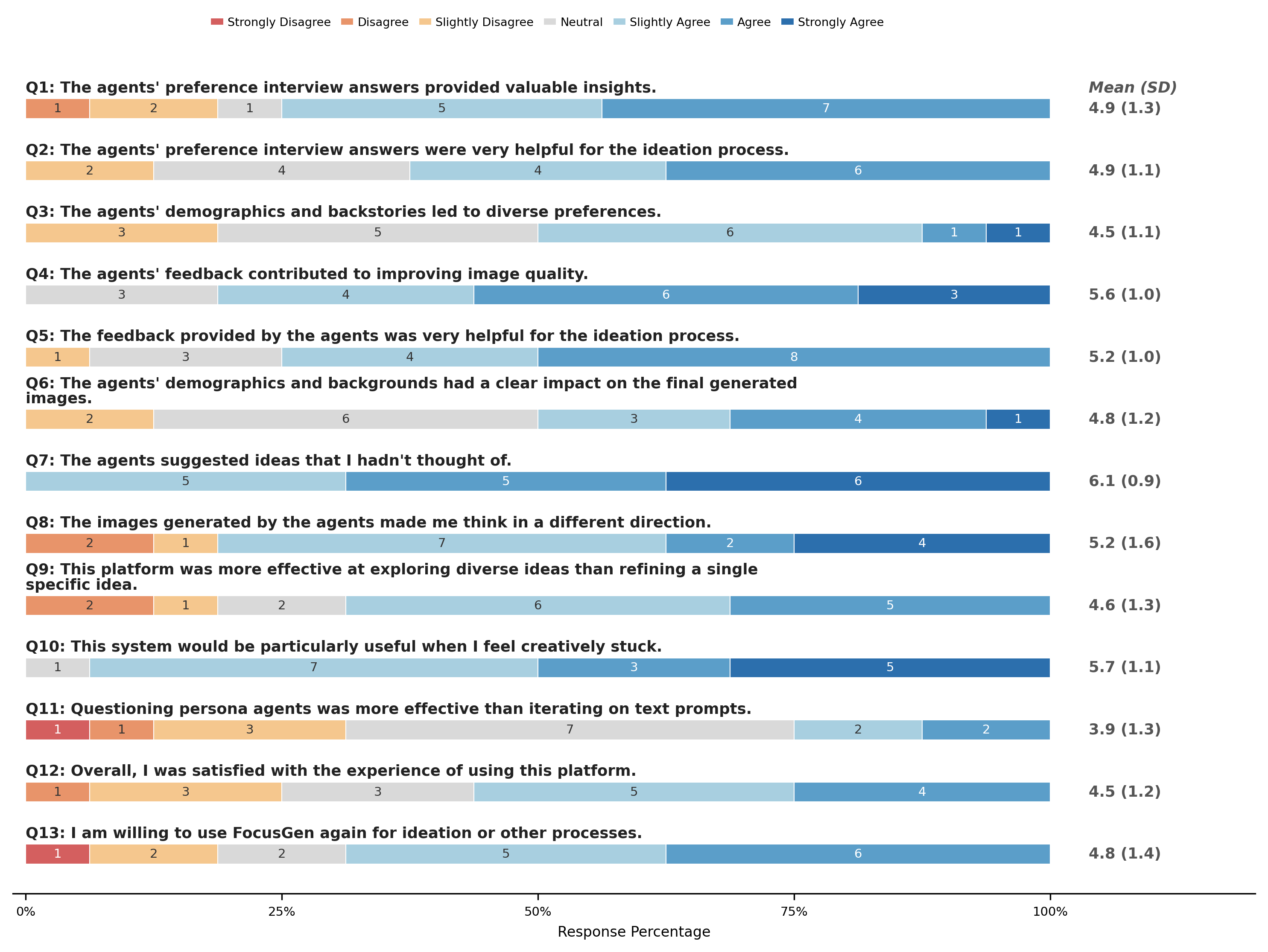}
    \caption{{Overall usability survey questionnaires and results for our system {\appname}}. (Q7 and Q11 are exploratory items; see Section 5.3.)}
    \Description{A chart visualizing the results of a 7-point Likert scale survey administered to 16 participants regarding their experience with FocusGen. The results show a high mean score for novelty (Q7, Mean=6.1) and usefulness when creatively stuck (Q10, Mean=5.7), moderately positive scores of approximately 4.5 to 5.6 for most other items, and a near-neutral score for the exploratory comparative item about prompting (Q11, Mean=3.9).}
    \label{fig:focusgen_user_survey_results}
\end{figure*}

\paragraph{Study Design Rationale.}
We used a fixed order (familiar tool first, {\appname} second) so that participants could anchor their reflections in a fresh, concrete experience of their established workflow. This design choice, together with the differing underlying models (participants' tools vs.\ Gemini/Imagen) and the difference in output volume (one image per generation vs.\ 20--30 in parallel), makes between-condition comparisons uninterpretable, and we do not make them. The Likert battery in Figure~\ref{fig:focusgen_user_survey_results} is reported as a description of participants' absolute experience with {\appname}. Two items require additional caution and should be read as exploratory. Q11 is comparative by construction (``\ldots more effective than iterating on text prompts'') and inherits all of the confounds above; its near-neutral mean ($M = 3.9$) should not be read as evidence in either direction. Q7 (``The agents suggested ideas that I hadn't thought of'') closely mirrors the system's premise and is therefore vulnerable to demand characteristics; in our analysis we weight the behavioral and artifact evidence---the specific concepts participants adopted, and the follow-up interview questions they formulated---more heavily than this self-report. Comparative claims would require a counterbalanced or between-subjects design with a matched generation backend, which we leave to future work.

\paragraph{Analysis.}
Two researchers independently coded the interview transcripts and submitted concepts using an open coding methodology, then collaboratively grouped codes into the themes presented below.

\subsection{Results}
Our analysis of the sessions and interviews revealed how creative professionals experienced and utilized the persona-driven workflow. We present our findings thematically, supported by qualitative feedback and descriptive statistics about the participants' experience with {\appname}. Full survey results are shown in Figure~\ref{fig:focusgen_user_survey_results}. We emphasize that the Likert items assess participants' experience with {\appname} in absolute terms; due to the confounds described above, we do not compare scores across conditions.

\subsubsection{{\appname} Unifies Visual Ideation in a Single Workflow}
Addressing \textbf{RQ1}, our first finding relates to how participants experienced the {\appname} workflow. When reflecting on their general practice, participants described their typical ideation process as fractured---many avoided generative tools altogether, preferring to browse visual-first sites like Pinterest. Those who did use AI described a cumbersome, two-stage process: first using an LLM for text ideas, then translating those into image prompts.

Participants found {\appname}'s single-workflow approach appealing after a brief adjustment period. While initially unfamiliar with asking preference questions instead of issuing direct commands, they quickly adapted. They appreciated receiving 20--30 unique visual concepts at once---a property of the system's parallel agent architecture---which mirrored the experience of browsing a mood board. As participant P12 noted, this meant they received \textit{``a whole board of visual starting points at once.''} Participant P5 similarly emphasized that \textit{``by setting just a broad direction at the beginning, I could receive results generated from various perspectives all at once.''}

\subsubsection{{\appname} Builds Confidence by Simulating Audience Feedback}
Further answering \textbf{RQ1}, we found that {\appname} builds designers' confidence, especially in unfamiliar domains {(Q10: M=5.7, SD=1.1)}. Many participants used their second session with {\appname} to validate the ideas they had developed in the first. For example, participant P1, a brand designer working on a clothing task, used the platform to check if his initial concepts aligned with the preferences of a simulated target audience. After seeing agents independently generate concepts similar to his own initial ideas (Figure~\ref{fig:participant_P1_images} in the Appendix), participant P1 felt significantly more confident and used that confidence as a basis for further, more adventurous exploration. In another example, P10, during the clothing task, asked step-by-step questions about text placement, colors, materials, and fit, receiving consistent preferences from AI agents (dark tones, long sleeves) along with specific suggestions (detachable hood, additional pockets). Participant P10 described this as \textit{``a good way for designers to understand how people think without directly conducting user research,''} emphasizing the value of rapid validation. In these cases, {\appname} served as a ``sanity check,'' allowing designers to quickly test their intuition against a simulated audience.

\subsubsection{Personas Generate Novel, Contextually-Rich Ideas}
Our central finding, speaking to both \textbf{RQ1} and \textbf{RQ2}, is how personas act as a catalyst for novel ideas that push users in new directions by providing diverse perspectives {(Q7: M=6.1, SD=0.9; Q9: M=4.6, SD=1.3)}. A clear example was the ``Coffee Brew 2025'' coffee festival poster task. In the baseline condition, participants used web searches or direct prompting, resulting in more conventional designs. With {\appname}, the results were far more diverse and contextually-aware (as seen in Figure~\ref{fig:genai_vs_focusgen} in the Appendix). Participant P6 was particularly impressed by a poster featuring a caffeine molecule---a completely different approach that she had not thought of. Participant P9 highlighted a concept where a drop of coffee fell onto a famous city landmark, an idea she found unique and clever.

Crucially, this novelty was not just visual; it was contextual. Participants reported that the agents' preference interview answers provided valuable insights (Q1: M=4.9, SD=1.3; Q2: M=4.9, SD=1.1). Unlike static images from Pinterest, these textual responses explained the ``why'' behind the ``what,'' serving a role similar to a docent explaining the intent behind a piece of art. Participant P11 described it: \textit{``For Assignment 2 [{\appname}], I felt like I engaged more in critical thinking, trying to understand the rationale behind the designs that the agents generated.''} Participant P16 similarly noted that interviewing agents enabled deeper reflection and thinking compared to simple prompting for image generation. This contextual understanding was generative in itself, often prompting participants to formulate new interview questions to explore these unexpected creative directions further.

\subsubsection{Demographics as a Creative Lever for Conceptual Exploration}
Providing an answer to \textbf{RQ2}, our study revealed how participants used demographic filters as a creative lever to probe the design space in unexpected ways. While the perceived impact of demographics was moderately significant overall {(Q6: M=4.8, SD=1.2)}, certain tasks provided salient examples of their influence.

The most striking instance occurred during the ``Beauty Expo 2025'' mascot task. Participants P9 and P12 initially sampled agents from one demographic. When they broadened their exploration to include another, the results suddenly included robot characters---a concept that had not appeared before (see Figure~\ref{fig:diff_gender_results} in the Appendix). Participant P12 found this revelatory: \textit{``It was really interesting... when I sampled agents that were [Gender B], some characters came out to be robots, which never appeared when I set the agents to all [Gender A].''} This discovery illustrates both the potential and the risk of demographic conditioning. On one hand, changing a single filter surfaced an entirely new conceptual direction that no participant had considered---exactly the kind of ``blind spot'' {\appname} is designed to reveal. On the other hand, the association between a specific demographic and a specific concept (robots in a beauty context) may reflect stereotypical patterns in the underlying model rather than genuine audience preferences. This tension is inherent to persona-based systems and underscores the importance of the designer's curatorial role: the value lies not in accepting any single agent's output as representative, but in using the diversity of outputs as provocations that expand the design space. We discuss mitigation strategies in Section~\ref{sec:ethics}.

This principle of contextual relevance extended beyond gender to agents' professional backgrounds. For the fintech logo task, participant P16 noted how a healthcare worker agent logically introduced medical-inspired design elements. Similarly, participant P9 found that the novel idea of using a caffeine molecule for a coffee poster \textit{``made sense''} because it was generated by an agent with a background in engineering. These instances show how an agent's specific life experience can generate logical yet unexpected conceptual blends that participants found valuable.

More broadly, participants valued demographic filters for grounding ideation in specific audience contexts, though this capability also raises questions about reinforcing stereotypes that we address in Section~\ref{sec:ethics}.

\subsubsection{Participants Positioned {\appname} as a Tool for Divergent Exploration}
Finally, answering \textbf{RQ2}, participants quickly formed a strong mental model of {\appname}'s ideal role. They agreed it would be most useful when creatively stuck {(Q10: M=5.7, SD=1.1)}, and they leaned toward viewing it as better suited to exploring diverse ideas than to refining a single concept {(Q9: M=4.6, SD=1.3; see also Q7: M=6.1, SD=0.9)}. Participant P6 evaluated {\appname} as \textit{``suitable for ideation while finding diverse references,''} emphasizing its utility for idea divergence. Participant P10 noted that \textit{``I could re-examine elements I took for granted and break existing preconceptions,''} recognizing {\appname}'s value as a tool for providing new perspectives. This positions {\appname} as a complementary, upstream tool whose primary value lies in breaking creative fixation at the earliest stages of a project.

\section{Discussion}
In this work, we introduced {\appname}, an interactive system that brings external audience perspectives into visual design exploration through a virtual focus group of persona agents. Our evaluations provide converging evidence that persona conditioning expands the diversity of generated design alternatives beyond what a generic assistant produces, and our qualitative study illustrates how creative professionals put that expanded spectrum to work---while also delineating what we have not shown: that agent outputs are discriminably faithful to their specific personas, or that they represent the preferences of real demographic groups (Sections 4 and 6.4).

\subsection{Ethical Considerations: Stereotyping and the Appeal of the Characterful}
\label{sec:ethics}

The demographic grounding that enables {\appname} to simulate diversity also risks reproducing demographic stereotypes~\cite{li_actions_2025, weidinger_taxonomy_2022}. Narrative backstories and preference interviews shift agents from demographic shorthand toward individualized traits~\cite{park_generative_2024}, but unlike Park et al.'s agents, which are grounded in interviews with real people, our backstories are themselves LLM-generated from demographic seeds. The mitigation is therefore partial and unaudited: the backstory-generation process may launder stereotypical associations into fluent individual narratives, and we did not conduct a formal bias audit.

Our own findings sharpen this concern rather than soften it. In the Beauty Expo case (Section 5.4.4), changing a single gender filter produced robot mascots, and participants found the outputs delightful and adopted the direction. Designer curation---which an earlier version of this paper offered as the primary safeguard---operated here in the opposite direction: the surprisingness that makes an output feel like a discovered blind spot is also what makes a stereotyped association attractive and likely to be incorporated. In a tool built to reward surprise, novelty-seeking and stereotype-adoption can be behaviorally indistinguishable. Similarly, P10 described the system as ``a good way for designers to understand how people think without directly conducting user research''---a reading our validation does not license, since we have not shown that agent preferences track those of real demographic groups.

We see three concrete obligations for systems like {\appname}. First, provenance and friction: the interface already attaches each image to the generating agent's full profile; it should additionally surface when a visual direction correlates strongly with a single demographic attribute across the cohort, prompting the designer to test that direction against other segments before adopting it. Second, framing: onboarding and documentation must present outputs as provocations for exploration, not as evidence of audience preference, since our participants did not always maintain this distinction unprompted. Third, auditing and validation: deployment beyond a research prototype requires a formal bias audit of the backstory- and preference-generation pipeline and validation of persona outputs against real demographic samples (Section 6.4). Until then, the appropriate claim for {\appname} is that it diversifies a designer's option set---not that it represents an audience.

\subsection{The Designer's Role: Agency and Ownership}
By automating the generation of visual concepts from multiple persona perspectives, {\appname} performs substantial creative labor with minimal real-time human input, raising questions about creative agency~\cite{frich_mapping_2019}. We position {\appname} deliberately as an \textit{upstream} ideation tool: the system generates a landscape of starting points, and the designer's expertise enters in curating, interpreting, and building upon these seeds. Our user study supports this---participants described the system as most valuable for early exploration, and P11 reported engaging in \textit{more} critical thinking because interpreting diverse agent perspectives demanded active judgment. The interview question design step, where the designer's creative intent shapes the system's behavior, may be the key moment of \textit{appropriation}~\cite{dix_designing_2007}. Future work should investigate how extended use affects designers' sense of ownership.

\subsection{Implications for Creativity Support Tool Design}

\subsubsection{From Dyadic to Polyphonic Co-Creation}
Current Human-AI interaction is predominantly dyadic. {\appname} proposes a shift toward \textbf{polyphonic} co-creation, extending crowd-powered design~\cite{andolina_crowdboard_2017, koyama_crowd-powered_2014} into the ``in silico'' domain, where interfaces present \textit{conflicting} perspectives simultaneously rather than optimizing for a single response~\cite{bodker_when_2006}.

\subsubsection{Narrative-Anchored Divergence over Stochastic Variation}
Our results suggest that \textit{contextual novelty} is far more valuable than stochastic variation~\cite{sadat_cads_2024, corso_particle_2023}. Persona-driven deviations provide a \textbf{semantic anchor}---a narrative logic that makes novelty plausible (e.g., an engineering agent suggesting a caffeine molecule). This extends work on meaningful diversity~\cite{han_poet_2025}: narrative grounding transforms diversity from noise into actionable inspiration.

\subsubsection{AI as a Mirror for Creative Self-Efficacy}
Participants used the system not just for inspiration but for internal validation---seeing intuitions reflected by a simulated audience gave them confidence to take risks~\cite{bandura_self-efficacy_1977}. This suggests AI simulation can function as a ``safe-to-fail'' sandbox~\cite{snowden_cynefin_2007} for creative risk-taking.

\subsubsection{Eliciting Dimensions versus Imposing Them}
Our open-ended-versus-structured finding connects directly to the dimension--value framework of Proxona~\cite{choi_proxona_2025}, in which audience personas are organized along creator-curated dimensions, and to dimensional-reasoning tools more broadly~\cite{cai_designaid_2023, suh_luminate_2024, tao_designweaver_2025}. Structured interviews in {\appname} function like designer-imposed dimensions: they guarantee coverage of axes the designer already considers relevant, which supports systematic comparison, but in our data they compress output diversity. Open-ended interviews instead let each persona nominate its own dimensions, trading comparability for breadth---and it is precisely this self-definition that appears to surface dimensions the designer had not considered. The two modes are complementary: a practical workflow might open with open-ended elicitation to map the space, then switch to structured questions to compare cohorts along the dimensions that emerged. This also clarifies our relationship to convergent persona systems such as PosterMate~\cite{shin_postermate_2025} and PersonaFlow~\cite{liu_personaflow_2025}: those systems concentrate persona perspectives onto one artifact or one research question, which suits refinement; the parallel-divergent topology suits the earlier moment when the designer does not yet know which artifact to refine.

\subsection{Limitations and Future Work}

\subsubsection{Synthetic Nature of Personas and the Missing Discriminative Test}
Our synthetic agents remain models without lived experience, and two validation gaps follow. First, discriminative faithfulness: our Technical Validation shows that the refinement loop improves alignment with a given preference profile, but we did not test whether outputs are discriminably tied to their specific persona---for example, whether participants prefer images generated from their own profile over images generated from another participant's profile, or whether persona-conditioned refinement outperforms refinement by a randomly assigned persona or a generic critic on the same profiles. This test is the right next step and is inexpensive, since it reuses the existing pipeline. Second, representativeness: we did not validate that synthetic agents' self-generated preferences match those of the demographic groups they are sampled from, and it is unknown to what extent the underlying models have learned genuine correlations between demographics and aesthetic preference (the illustrative carpenter example in Section 3.2.2 should be read as a description of the mechanism, not as a validated association). A valuable community resource here would be a calibration dataset of real aesthetic preferences across demographic strata, against which persona pipelines could be re-validated as underlying models change; absent such a dataset, results like ours are tied to specific model versions (gemini-2.5-flash, imagen-3.0-generate) and must be re-established when models change. Future work should also ground personas in real, anonymized interview data~\cite{park_generative_2024} and conduct systematic bias audits of the backstory-generation process.

\subsubsection{Non-Monotonic Refinement}
The refinement process is non-monotonic---later iterations do not always improve upon earlier ones. Our design choice to have agents review the \textit{entire} image sequence and select the best (rather than defaulting to the last) addresses this, but future work should investigate more principled convergence criteria.

\subsubsection{Generalizability}
Our study was limited to 16 creative professionals from a non-U.S. country, and our Agent Bank uses U.S. census data. This cultural mismatch may inflate perceived novelty---some concepts may have felt surprising due to cultural distance rather than the persona mechanism itself. Future work should match agent demographics to the designer's target market, replicate with broader participant samples, diverse cultural datasets, and extend beyond static images to domains like prototyping and video.

\section{Conclusion}
Creative professionals design for audiences, not just themselves---yet current text-to-image exploration tools derive diversity entirely from the designer's own input, limiting discovery to what the designer already knows to look for. In this paper, we introduced {\appname}, an interactive system that brings external perspectives into visual design exploration through a virtual focus group of persona agents. Each agent, constructed from demographic data, a procedural backstory, and elicited aesthetic preferences, independently generates visual concepts conditioned on that profile---surfacing directions the designer would not have explored alone. Our evaluations showed that agent diversity translates directly into visual diversity, and that open-ended preference interviews expand the design space by letting agents define their own aesthetic dimensions. A study with 16 creative professionals found that {\appname} helped them discover unanticipated directions---from caffeine molecules on coffee posters to robots in beauty mascots---overcome fixation, and build confidence through simulated audience feedback.

These perspectives are simulations: our evaluations show that they expand the diversity of a designer's option set and that designers find the expansion useful, while leaving open---and we have made explicit---the question of whether they are faithful to the specific people and groups they are modeled on.

This work contributes to a broader shift in human-AI creative partnership: from engineering prompts for a single AI toward curating a diverse, persona-driven ``virtual focus group'' that surfaces the blind spots designers need to see.

\bibliographystyle{ACM-Reference-Format}
\bibliography{references, references_extra}

\newpage
\appendix

\begin{figure*}[tb]
    \centering 

    \begin{subfigure}{0.4\textwidth} 
        \centering
        \includegraphics[width=\linewidth]{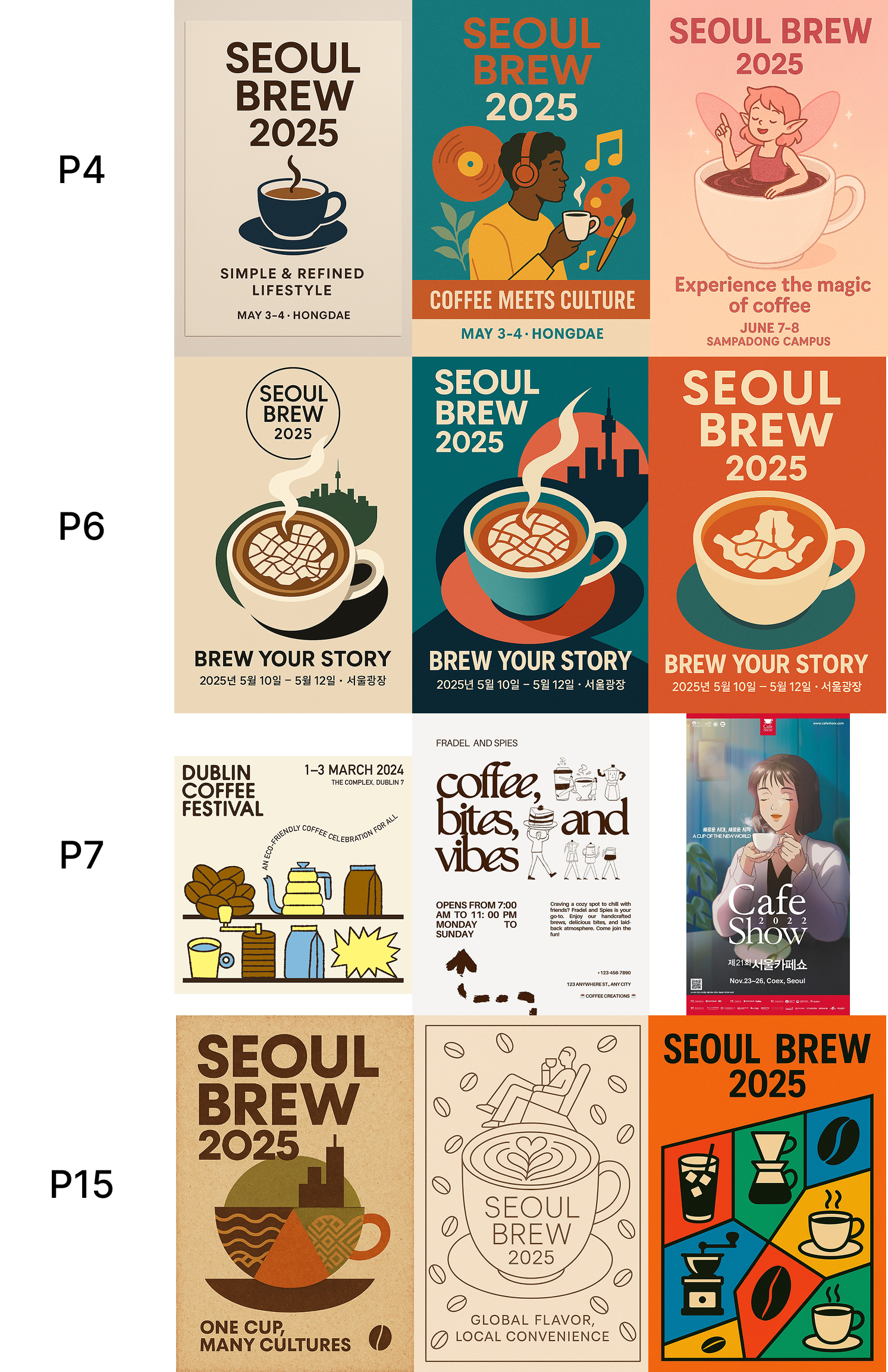}
        \subcaption{Visual concept from a baseline tool.}
        \label{fig:genai_vs_focusgen_a}
    \end{subfigure}
    \hfill
    \begin{subfigure}{0.48\textwidth} 
        \centering
        \includegraphics[width=\linewidth]{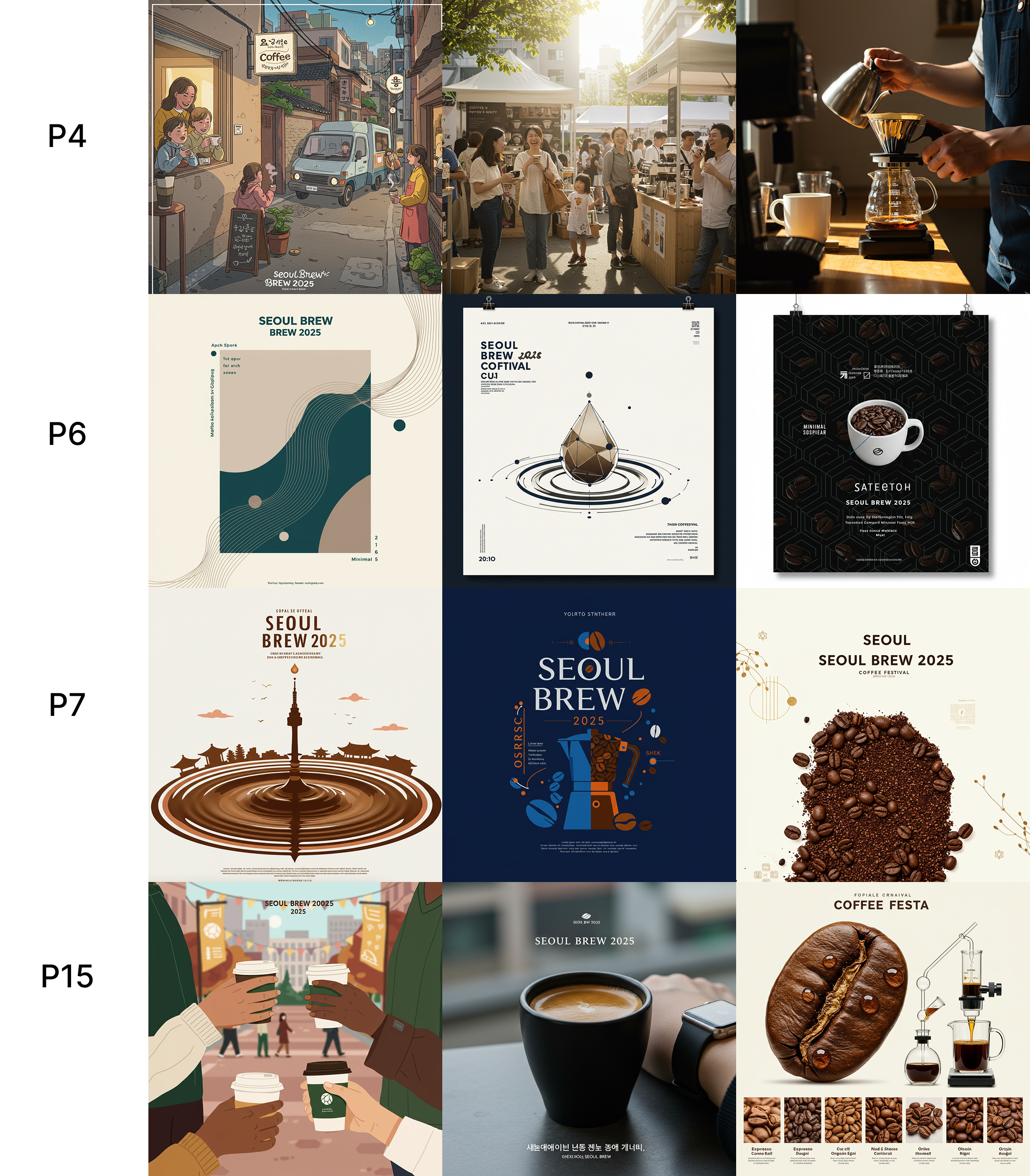}
        \subcaption{Visual concept from {\appname}.}
        \label{fig:genai_vs_focusgen_b}
    \end{subfigure}

    \caption{Comparison of concepts generated for the ``Coffee Brew 2025'' poster task. Concepts from baseline tools (left) were often conventional, while {\appname} (right) generated more diverse and contextually-aware ideas, such as the caffeine molecule and the coffee drop on the city landmark, which participants found novel.}
    \Description{A two-panel figure comparing poster designs for a 'Coffee Brew 2025' coffee festival. The left panel, from a baseline tool, displays 12 posters in a similar, conventional illustrative style featuring coffee cups. The right panel, from FocusGen, displays 12 posters in a wide variety of styles, including photorealism and abstract graphics. The FocusGen concepts are more diverse, featuring novel ideas like a city landmark made of coffee and a representation of a caffeine molecule.}
    \label{fig:genai_vs_focusgen}
\end{figure*}

\begin{figure*}[h!]
    \centering 

    \begin{subfigure}{0.8\textwidth}
        \centering
        \includegraphics[width=\linewidth]{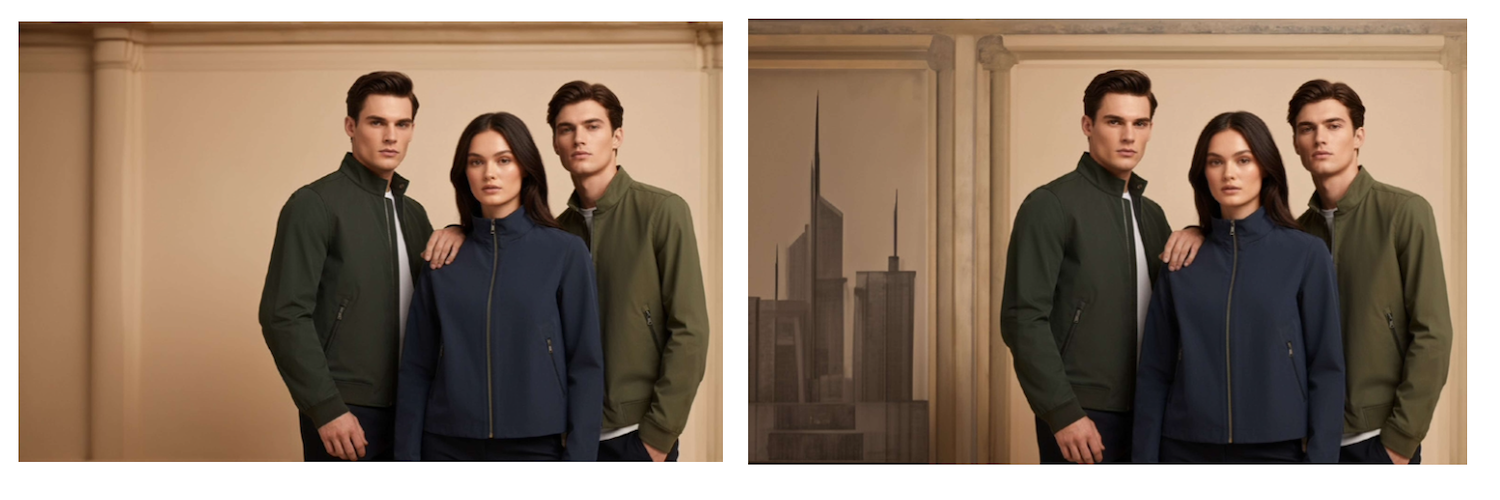}
        \subcaption{Images generated by Participant P1 using his preferred AI image tool.}
        \label{fig:participant_P1_images_a}
    \end{subfigure}
    \hfill 

    \begin{subfigure}{0.8\textwidth}
        \centering
        \includegraphics[width=\linewidth]{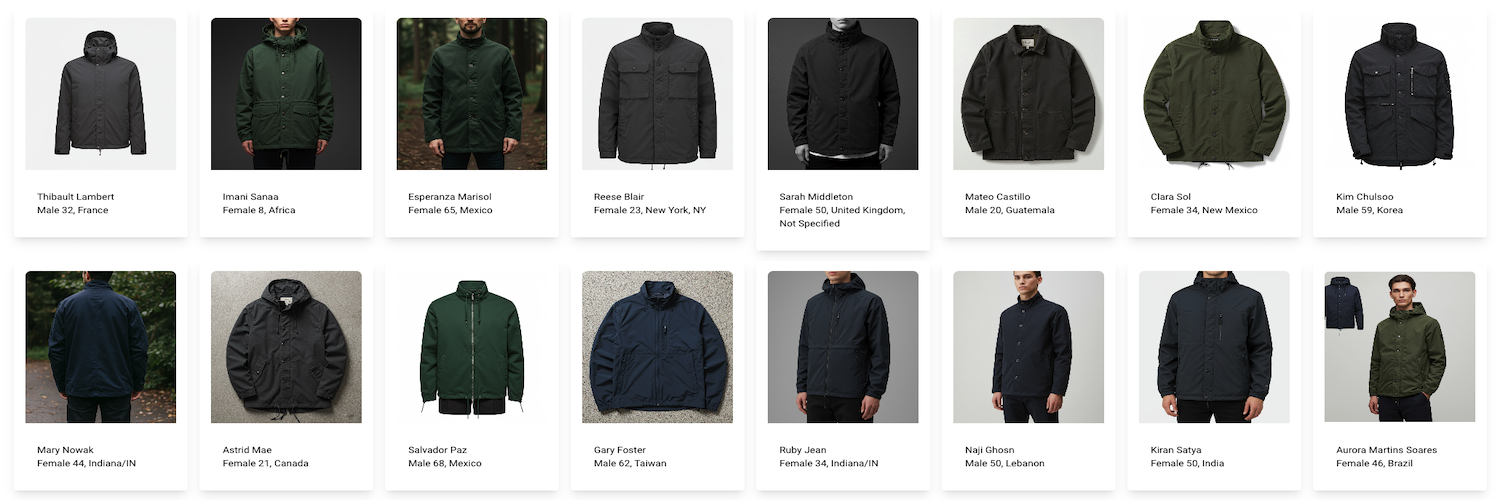}
        \subcaption{First batch of images generated by Participant P1 using {\appname}.}
        \label{fig:participant_P1_images_b}
    \end{subfigure}
    \hfill 

    \begin{subfigure}{0.8\textwidth}
        \centering
        \includegraphics[width=\linewidth]{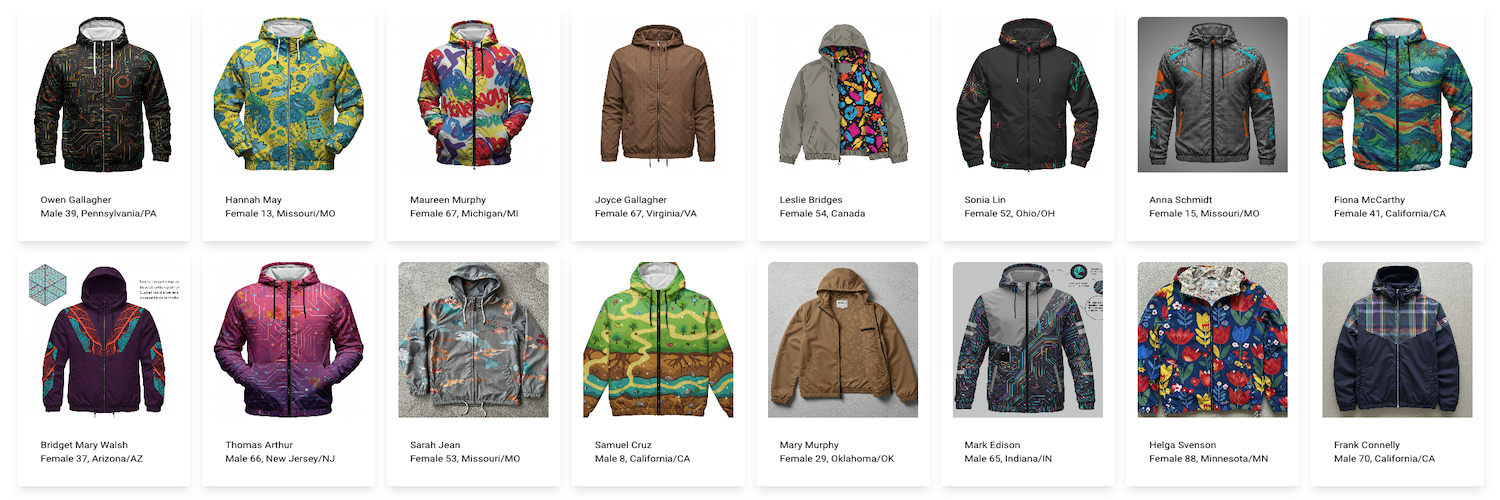}
        \subcaption{Subsequent batch of images generated by Participant P1 using {\appname}.}
        \label{fig:participant_P1_images_c}
    \end{subfigure}

    \caption{An example of how participant P1 used {\appname} to build creative confidence: P1's initial concepts, generated with a baseline tool as shown in~\ref{fig:participant_P1_images_a}, were validated when {\appname}'s agents independently produced similar ideas as shown in~\ref{fig:participant_P1_images_b}. This confirmation gave participant P1 the confidence to then use the system to explore more adventurous, novel concepts (in~\ref{fig:participant_P1_images_c}).}
    \Description{A three-panel figure illustrating a participant's creative workflow. Subfigure (a) shows the participant's initial concepts generated using participant's preferred AI tool: models in simple, solid-colored jackets. Subfigure (b) shows a grid of stylistically similar jackets generated by FocusGen, which validated the idea. Subfigure (c) shows a subsequent grid of jackets from FocusGen that are much more adventurous, featuring bright, multi-colored patterns. The figure demonstrates how initial validation can lead to more creative exploration.}
    \label{fig:participant_P1_images}
\end{figure*}

\begin{table*}[htb]
\centering
\renewcommand{\arraystretch}{1.3} 
\begin{tabular}{p{0.2\textwidth} p{0.6\textwidth} p{0.1\textwidth}}
\toprule
\textbf{Input Prompt Type} & \textbf{Prompt Content} & \textbf{Token Count} \\
\midrule
Demographic Profile &
    \begin{minipage}[t]{\linewidth}
    \vspace{-\baselineskip} 
    \begin{itemize}
        \item Name: Mia
        \item Age: 29
        \item Gender: Female
        \item Race: White alone
        \item Education: graduate degree
        \item ...
        \item Place of Birth: New Mexico/NM
    \end{itemize}
    \end{minipage}
    & $\approx$ 150 \\
\midrule
Backstory &
    I was born 29 years ago in New Mexico, a state that always felt vast and quiet, painted with colors of red earth and infinite sky. My childhood wasn't particularly eventful, at least not in the dramatic sense. It was... Growing up in New Mexico was a calm existence. We...
    My focus was always on my studies. High school was a period of intense academic application. I thrived on...
    When it came time for college, I moved away from New Mexico, drawn by a program that offered a rigorous academic curriculum. I pursued a Bachelor's degree, followed quickly by a Master's degree in business, with a specialization in risk management and finance...
    It was during my Master's program that I met my husband, Alex... After graduating, I went straight into the insurance industry... I spend my days sifting through information, scrutinizing facts, determining coverage and liability...
& $\approx$ 1000 \\
\midrule
Preference Profile &
    \begin{minipage}[t]{\linewidth}
    Here are your interview responses for an ideal bedroom:
    \begin{itemize}
        \item I would use matte white paint for the walls with one accent wall in muted sage green. The floor would...
        \item I want a low-profile platform bed with a walnut wood frame. The bedding would be...
        \item I'd include a small desk in the corner near the window, a full-length mirror...
    \end{itemize}
    \end{minipage} & $\approx$ 400 \\
\midrule
    Instruction Prompt & Your task is to roleplay as this character and answer questions based on your profile. Your responses should reflect your character's personality, background, and experiences. & $\approx$ 200 \\
\bottomrule
\end{tabular}
\caption{The prompt for a user-simulating persona agent is constructed by concatenating the user's demographic profile and personal backstory, their elicited preference interview data, and a final instruction for the agent to adopt this profile.}
\label{tab:agent-prompts}
\Description{A three-column table showing the structure of a prompt for a user-simulating persona agent. The columns are 'Input Prompt Type,' 'Prompt Content,' and 'Token Count.' The table has four rows detailing the prompt's components: a 'Demographic Profile' with example demographic profile (e.g., Name, Age, Gender, Race, Education, Place of Birth) and around 150 expected token count, a 'Backstory' with an example backstory of the agent (e.g., I was born 29 years ago in New Mexico, a state that always felt vast and quiet, painted with colors of red earth and infinite sky. My childhood wasn't...) and about 1000 expected token count, 'Task Preference Data' with example aesthetic preferences with around 400 expected token count, and a final 'Instruction Prompt' commanding the AI to roleplay with expected token count of 200.}
\end{table*}

\begin{table*}[tb!]
\centering
\newcolumntype{L}{>{\raggedright\arraybackslash}X}
\setlength{\tabcolsep}{6pt} 
\begin{tabularx}{\textwidth}{@{} llrll @{}}
\toprule
\textbf{ID (Gender, Age)} & \textbf{Profession/Major} & \textbf{YoE} & \textbf{Industry/Major (Employment)} & \textbf{Preferred Gen AI Tools*} \\
\midrule
P1 (M, 30s)  & Brand Designer      & 8  & Finance (Full-time) & \textbf{Adobe AI}, Shutterstock AI \\
P2 (F, 20s)  & Art Director        & 3  & Ad Agency H (Full-time) & \textbf{ChatGPT}, Midjourney \\
P3 (F, 30s)  & Art Director        & 10 & Ad Agency H (Full-time) & \textbf{ChatGPT} \\
P4 (F, 30s)  & Art Director        & 10 & Ad Agency D (Full-time) & \textbf{ChatGPT}, Midjourney \\
P5 (F, 30s)  & Brand Designer      & 11 & Ad Agency C (Full-time) & \textbf{ChatGPT}, Adobe AI, Midjourney \\
P6 (F, 30s)  & Art Director        & 10 &  Ad Agency H (Full-time) & \textbf{ChatGPT}, Midjourney, Adobe AI \\
P7 (F, 30s)  & Art Director        & 8  & Ad Agency H (Full-time) & \textbf{ChatGPT, Gemini} \\
P8 (F, 30s)  & Fashion(Jewelry) Designer    & 3  & Startup (Full-time) & \textbf{ChatGPT, Midjourney} \\
P9 (F, 30s)  & Art Director        & 10 & Ad Agency H (Full-time) & \textbf{ChatGPT} \\
P10 (F, 20s) & Master's Student    & 4  & Industrial Design (Full-time) & \textbf{ChatGPT, Midjourney} \\
P11 (M, 20s) & Master's Student    & 3  & Industrial Design (Full-time) & \textbf{ChatGPT, Midjourney} \\
P12 (F, 20s) & Freelancer          & 4 & Self-Employed (Part-time) & \textbf{ChatGPT} \\
P13 (F, 30s) & Art Director        & 8 & Ad Agency H (Full-time) & \textbf{ChatGPT} \\
P14 (F, 30s) & Freelancer          & 10 & Self-Employed (Part-time) & \textbf{ChatGPT}, Midjourney \\
P15 (F, 30s) & Art Director        & 8 & Ad Agency H (Full-time) & \textbf{ChatGPT}, Veo3 \\
P16 (M, 20s) & UX Designer         & 4  & Education Tech (Full-time) & \textbf{ChatGPT}, Bolt.new, DALL-E \\
\bottomrule
\end{tabularx}
\caption{Summary of the 16 creative professionals who participated in our user study. The table details their demographic and professional backgrounds, years of experience, and preferred generative AI tools. The tool indicated in \textbf{bold} is the one each participant chose to use for the baseline condition in our study.}
\Description{A table summarizing demographic and professional information for the 16 participants in the user study. The table has five columns: ID (including gender and age), Profession/Major, Years of Experience, Industry/Major, and Preferred Gen AI Tools. The participants are primarily creative professionals such as Art Directors and Brand Designers, with experience ranging from 3 to 11 years. The most frequently cited AI tool is ChatGPT. The caption clarifies that the bolded tool in the list is the one each participant used for the study's baseline condition.}
\label{tab:participants_info}
\end{table*}

\begin{table*}[tb!]
\centering
\caption{An overview of the six design ideation tasks used in our user study. The tasks were designed to cover a diverse range of creative scenarios, each with a unique brief and set of constraints.}
\label{tab:design_tasks_overview}
\begin{tabularx}{\textwidth}{clX}
\toprule
\textbf{No.} & \textbf{Task Name} & \textbf{Task Overview} \\
\midrule
1 & Logo Design Ideation & Collect logo references for a new fintech service, \textbf{``Future Pay''}. The core values are \textbf{``Transparency, Speed, and Trust''}. \\
\midrule
2 & Character Design Ideation & Collect references for the mascot of the \textbf{``Beauty Expo 2025''}. The character will be used to expand the brand's tone on packaging, app icons, and social media. \\
\midrule
3 & Clothing/Shoe Design Ideation & Collect outerwear references for the \textbf{``Urban Run 2025''} sports festival's capsule collection. The main item is a \textbf{unisex hybrid jacket} for both urban commuting and outdoor activities. \\
\midrule
4 & Ad Poster Design Ideation & Collect poster references for the \textbf{``Coffee Brew 2025''} coffee festival. The poster should emphasize \textbf{``simplicity + a stylish lifestyle''} and will be used in cafes, at bus stops, and on social media. \\
\midrule
5 & Social Media Campaign Ideation & Collect visual references for the \textbf{``Summer Spark 2025''} social media campaign. Ads will run on Instagram and TikTok, needing a mood that evokes \textbf{``coolness, trendiness, and a desire to share''}. \\
\midrule
6 & Freestyle Design Ideation & Assume you receive an \textbf{``open brief''} from a new client entering a new market. The client wants to visually explore their brand's tone and mood. \\
\bottomrule
\end{tabularx}
\Description{A three-column table outlining the six design ideation tasks used in the study. The columns are 'No.,' 'Task Name,' and 'Task Overview.' The six tasks are: 1. Logo Design for a fintech service, 2. Character Design for a beauty expo, 3. Clothing Design for a sports festival jacket, 4. Ad Poster Design for a coffee festival, 5. Social Media Campaign visuals, and 6. a Freestyle Design task with an open brief.}
\end{table*}

\begin{figure*}[tb!]
    \centering
    \includegraphics[width=1.0\linewidth]{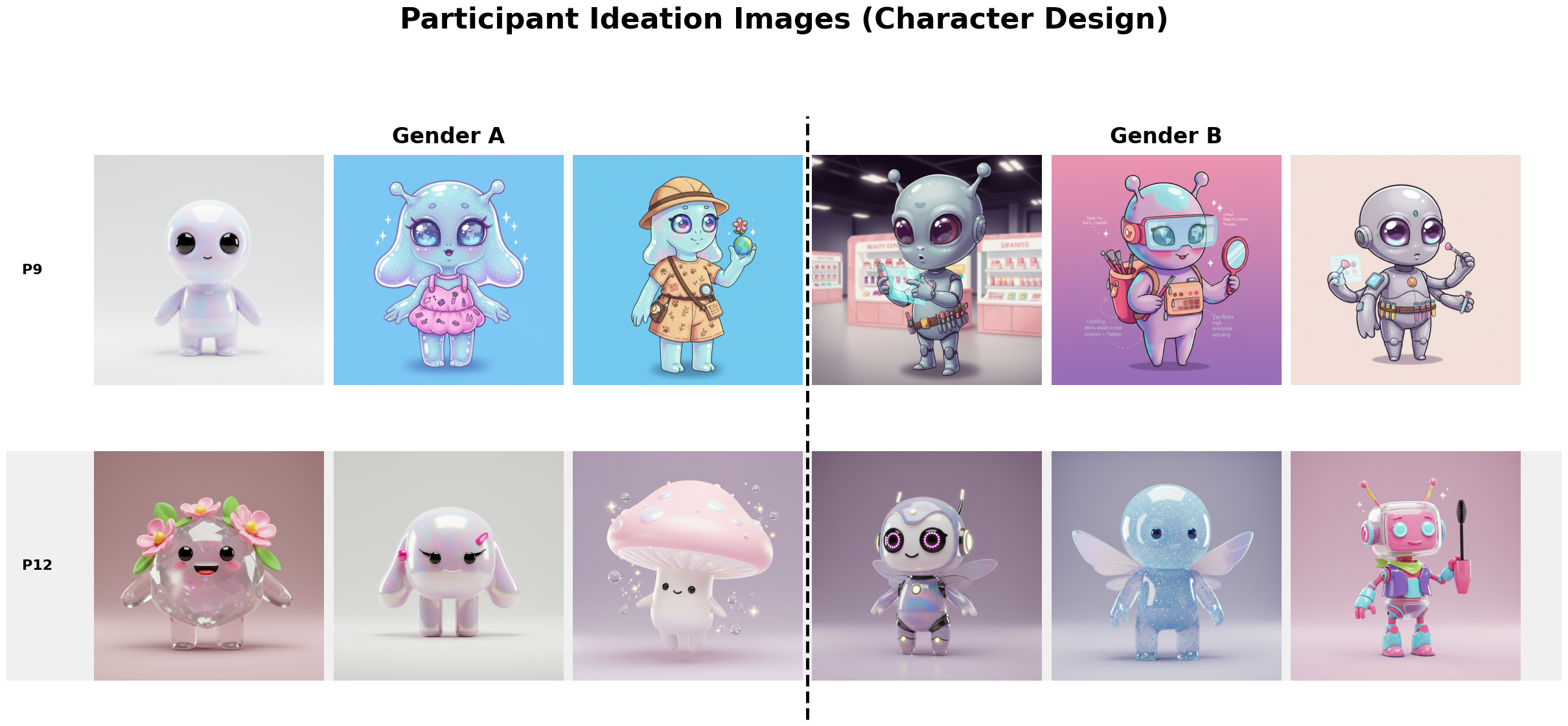}
    \caption{A salient example of demographic filters influencing conceptual direction from the ``Beauty Expo 2025'' mascot task. \textbf{Left:} Images from agents sampled from Gender A only. \textbf{Right:} Images from agents sampled from all genders.}
    \Description{A two-panel figure showing character designs for a 'Beauty Expo 2025' mascot from two participants, P9 and P12. The left panel, labeled 'Gender A,' displays six cute, organic-style characters, including aliens and a sentient mushroom. The right panel, labeled 'Gender B,' displays six characters that incorporate a robotic or technological theme, such as a robot bee and an alien with multiple robotic arms. The figure illustrates how changing the gender filter introduced the concept of robotics into the designs.}
    \label{fig:diff_gender_results}
\end{figure*}

\begin{figure*}[tb!]
    \centering
    \includegraphics[width=1.0\linewidth]{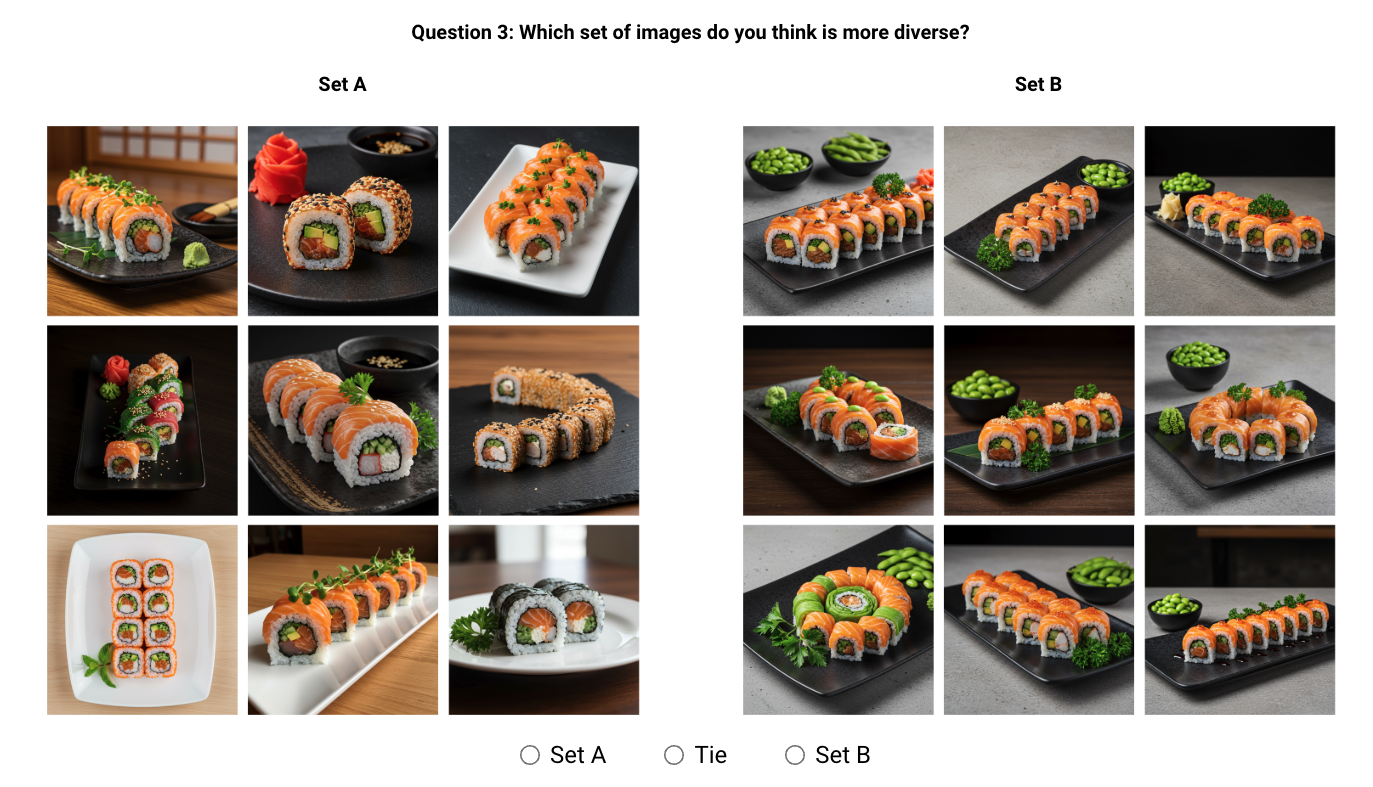}
    \caption{The interface for the comparative study in our diversity evaluation. Participants were shown two 3x3 grids of images for a given creative task and were asked to select which grid exhibited greater visual diversity, or to choose a tie. For each pair, one grid was generated using the Persona-Disabled (Baseline) condition, while the other was generated using the full, Persona-Enabled {\appname} system.}
    \Description{A screenshot of the user interface for the diversity evaluation study. The interface displays the question, "Which set of images do you think is more diverse?". Below the question are two 3x3 grids of images labeled 'Set A' and 'Set B', both showing various photos of sushi. The images in Set A are visually diverse in composition and type, while the images in Set B are more visually similar to one another. At the bottom, three radio buttons allow the user to choose 'Set A', 'Tie', or 'Set B'.}
    \label{fig:persona_agent_evaluation_interface}
\end{figure*}

\end{document}